\documentclass[twocolumn]{aastex62}

\usepackage{epsfig}
\usepackage{epstopdf}
\usepackage{graphics}
\usepackage{amsmath}
\usepackage{xcolor}

\usepackage{array}
\usepackage{booktabs}
\usepackage{amsmath,amssymb,amsfonts,amsbsy}
\usepackage{mathrsfs}
\usepackage{url}
\usepackage{multirow}
\usepackage{xcolor}
\usepackage{color}
\usepackage{ulem}
\usepackage{enumerate}
\usepackage{lineno}

\renewcommand{\figurename}{Fig.}
\renewcommand{\tablename}{Tab.}
\begin{document}

\title{Understanding the phase reversals of Galactic cosmic ray anisotropies}

\correspondingauthor{ Qiang Yuan, Yi-Qing Guo}
\email{ yuanq@pmo.ac.cn, guoyq@ihep.ac.cn}

\author{Bing-Qiang Qiao}
\affiliation{Key Laboratory of Particle Astrophysics,
	Institute of High Energy Physics, Chinese Academy of Sciences, Beijing 100049, China
}

\author{Qing Luo}
\affiliation{Hebei Normal University, Shijiazhuang 050024 , Hebei, China
}
\affiliation{Key Laboratory of Particle Astrophysics,
	Institute of High Energy Physics, Chinese Academy of Sciences, Beijing 100049, China
}

\author{Qiang Yuan}
\affiliation{
	Key Laboratory of Dark Matter and Space Astronomy, Purple Mountain Observatory, Chinese Academy of Sciences, Nanjing 210023, China
}
\affiliation{
	School of Astronomy and Space Science, University of Science and Technology of China, Hefei 230026, China
}

\author{Yi-Qing Guo}
\affiliation{Key Laboratory of Particle Astrophysics,
	Institute of High Energy Physics, Chinese Academy of Sciences, Beijing 100049, China
}
\affiliation{University of Chinese Academy of Sciences, 19 A Yuquan Rd, Shijingshan District, Beijing 100049, China}

\begin{abstract}
{
The energy spectra and anisotropies are very important probes of the origin of cosmic rays. Recent measurements show that complicated but very interesting structures exist, at similar energies, in both the spectra and energy-dependent anisotropies, indicating a common origin of these structures. Particularly interesting phenomenon is that there
is a reversal of the phase of the dipole anisotropies, which challenges a theoretical modeling. In this work, for the first time, we identify that there might be an additional phase reversal at $\sim100$ GeV energies of the dipole anisotropies as indicated by a few underground muon detectors and the first direct measurement by the Fermi satellite, coincident with the hundreds of GV hardenings of the spectra. We propose that these two phase reversals, together with the energy-evolution of the amplitudes and spectra, can be naturally explained with a nearby source overlapping onto the diffuse background. As a consequence, the spectra and anisotropies can be understood as the scalar and vector components of this model, and the two reversals
of the phases characterize just the competition of the cosmic ray streamings between the nearby source and the background. The alignment of the cosmic ray streamings along the local large-scale magnetic field
may play an important but sub-dominant role in regulating the cosmic ray propagation. More precise measurements of the anisotropy evolution at both low energies by space detectors and high energies by air shower experiments for individual species will be essential to further test this scenario.
}
\end{abstract}

\section{Introduction}
\par
The origin of Galactic cosmic rays (GCRs), after more than one century of
their first discovery, remains one of the most interesting mysteries in
astrophysics. Directly identifying the sources of GCRs is difficult, since
those charged particles propagate diffusively in the turbulent magnetic
field of the Milky Way, losing largely their original directions. The
information of the sources and the propagation process can imprint in the
energy spectra and the tiny anisotropies of GCRs
\citep{2013A&ARv..21...70B,2015ARA&A..53..199G,2017PrPNP..94..184A}.
Precise measurements of the spectra and anisotropies of GCRs, particularly
for various species, are thus very important in probing the origin of GCRs.

There have been quite a lot of efforts in measuring the anisotropies of 
arrival directions of GCRs, at large and small scales, by underground and
ground-based experiments
\citep{2005ApJ...626L..29A,2007PhRvD..75f2003G,2009ApJ...698.2121A,
	2010ApJ...718L.194A,2012ApJ...746...33A,2013ApJ...765...55A,
	2015ApJ...809...90B,2017ApJ...836..153A,2018ApJ...861...93B,
	2006Sci...314..439A,2008PhRvL.101v1101A,2010ApJ...711..119A,
	2011ApJ...740...16A,2013PhRvD..88h2001B,2014ApJ...796..108A,
	2016ApJ...826..220A,2019ApJ...871...96A}.  
The dominant component of the anisotropies 
is approximately a dipole, with amplitudes being $10^{-4}\sim10^{-3}$ for 
energies below PeV and experiencing a complicated rising-declining-rising
energy-dependence. The phase of the dipole anisotropies shows a reversal at 
$\sim 100$ TeV when the amplitude reaches the minimum. For $E\lesssim 100$ 
TeV, the phase is aligned approximately with the local interstellar magnetic 
field as inferred by IBEX measurements of the neutral atoms
\citep{2014Sci...343..988S}, and at higher energies the phase turns to 
the Galactic center (GC) direction.

The explanation of the GCR anisotropies is a challenge for a long time.
The conventional GCR production and propagation model with diffusion
coefficients obtained from fitting the B/C ratio predicts in general
a power-law rising of the dipole amplitude which is significantly higher
than the measurements \citep{2007ARNPS..57..285S}. The predicted dipole
phase of the conventional model is from the GC to the anti-GC direction
which also differs from the data \citep{2007ARNPS..57..285S}.
The observed energy-dependence of the amplitude and phase thus triggers
a lot of discussion of possible effects from discrete sources and/or large 
scale magnetic fields \citep{2006APh....25..183E,2012JCAP...01..011B,
2013ApJ...766....4P,2013APh....50...33S,2014ApJ...785..129K,
2015PhRvL.114b1101M,2015ApJ...809L..23S,2016PhRvL.117o1103A}.
{For example, \citet{2015ApJ...809L..23S} illustrated that 
the dipole anisotropies of GCRs could be reproduced assuming a superposition
of a local source with an age of 2 Myr and a distance of $\sim$200 pc, 
and the contribution from the background GCR sea.}

Recently, new observations of the energy spectra of GCRs and $\gamma$
rays shed new light on the understanding of the anisotropies and further
on the origin of GCRs. Precise measurements of the energy spectra of
protons and helium nuclei show hardening features around several hundred
GV rigidities and subsequent softening features around $\sim10$ TV
\citep{2015PhRvL.114q1103A,2015PhRvL.115u1101A,2019SciA....5.3793A,
	2021PhRvL.126t1102A}. It is very interesting to note that the spectral
structures of the fluxes may co-evolve with the anisotropies, indicating
common origins of those structures \citep{2019JCAP...10..010L,
	2019JCAP...12..007Q}. In addition, the extended $\gamma$-ray morphologies
around middle-aged pulsars, known as pulsar halos, suggest that the
diffusion of high-energy particles in the surroundings of pulsars is
much slower than the average diffusion velocity in the Milky Way
\citep{2017Sci...358..911A,2021PhRvL.126x1103A}. The spatial variations
of the GCR intensities and spectral indices inferred from the Fermi
$\gamma$-ray observations also suggest the propagation of GCRs is
inhomogeneous throughout the Milky Way
\citep{2016PhRvD..93l3007Y,2016ApJS..223...26A}.
A spatially inhomogeneous propagation scenario plus a nearby source
was proposed, which naturally explains the new phenomena of GCR spectra
and $\gamma$ rays, as well as the anisotropies
\citep{2019JCAP...10..010L,2019JCAP...12..007Q}.

The key point of the above scenario to account for the phase reversal
of the dipole anisotropies at $\sim100$ TeV is the competition of the
GCR streamings between the background sources (from the GC to the anti-GC
direction, dominating above 100 TeV) and the nearby source (from the
source direction to the opposite direction, dominating below 100 TeV).
The projection of the streamings along the local magnetic field may
affect the exact values of the dipole phase \citep{2016PhRvL.117o1103A},
but the overall picture keeps applicable. A natural expectation of this
scenario is that there should be an additional phase reversal at low
energies when the background streaming dominates again since the
background GCR spectrum is softer than the local source. It is very
interesting to note that the first direct measurement of the all-sky
anisotropies by space satellite Fermi seems to observe this predicted
reversal of the dipole component at $\sim100$ GeV energies, although
the significance is not high enough to robustly claim a detection
\citep{2019ApJ...883...33A}. The Fermi data shows a dipole amplitude
of $(3.9\pm1.5)\times10^{-4}$ and a phase of $-9.7\pm1.5$ hr in the
Right Ascension (R.A.) when projecting to one-dimension on R.A., for
protons with $E>78$ GeV \citep{2019ApJ...883...33A}. The phase of Fermi
result differs by nearly 12 hr from those measured by underground muon
detectors above 100 GeV, and just experiences a reversal.


\section{Model Description}
\subsection{The propagation of GCRs} 
The propagation of GCRs in the Milky Way is naturally expected to be 
inhomogeneous due to the spatial variations of the physical parameters 
of the interstellar medium (ISM). The non-linear feedback of GCRs on 
the ISM turbulence may further enhance the spatial inhomogeneities of 
the GCR transportation \citep{2012PhRvL.109f1101B,2016MNRAS.462L..88R}. 
The observations of $\gamma$-rays by HAWC, and LHAASO offer direct 
evidence in support of such a scenario 
\citep{2017Sci...358..911A,2021PhRvL.126x1103A}.
In this work we adopt a two-halo description of the spatially-dependent 
propagation (SDP) of GCRs \citep{2012ApJ...752L..13T,2016ApJ...819...54G}.
Here, the two propagation halos are defined as the inner halo (disk) and 
outer halo. The diffusion in the inner halo is much slower than that in 
the outer one. The diffusion coefficient can be parameterized as
\begin{equation}
D_{xx}(r,z, {\cal R} )= D_{0}F(r,z)\beta^{\eta} \left(\dfrac{\cal R}
{{\cal R}_{0}} \right)^{\delta_0 F(r,z)},
\label{eq:diffusion}
\end{equation}
where ${\cal R}$ is the particle's rigidity, ${\cal R}_0\equiv 4$ GV, 
$D_0$ and $\delta_0$ are constants representing the diffusion coefficient 
in the outer halo, $\eta$ is a phenomenological constant employed to fit the 
low-energy spectra. The spatial-dependent function $F(r,z)$ is defined as
\citep{2016ApJ...819...54G},
\begin{equation}
F(r,z) = \left\{
\begin{array}{ll}
{g(r,z)+[1-g(r,z)]} \left(\dfrac{{z}}{\xi{z_h}} \right)^{n},  &  {{|z|} \leq \xi{z_h}} \\
1,  &  { {|z|} > \xi{z}_{\rm h}} \\
\end{array},
\right.
\end{equation}
where $g(r,z)$=$N_m$/[1+$f(r,z)$] with $f(r,z)$ being the GCR source 
distribution, $z_h$ is the half-thickness of the propagation cylinder, 
and $\xi z_{h}$ is the half-thickness of the inner halo. The factor 
$\left(\frac{{z}}{\xi{z}_{\rm h}} \right)^{n}$ enables a smooth 
transition of the diffusion parameters between the two halos. 
The spatial dependence of the diffusion coefficient is phenomenologically 
assumed, which shows a general anti-correlation with the source function. 
Physically it may be related with the magnetic field distribution, or the 
turbulence driven by GCRs \citep{2012PhRvL.109f1101B,2016MNRAS.462L..88R}. 
We adopt the diffusion re-acceleration framework in this work. 
The re-acceleration is described by the momentum diffusion with
\begin{equation}
D_{pp}D_{xx} = \dfrac{4p^{2}v_{A}^{2}}{3\delta(4-\delta^{2})(4-\delta)},
\end{equation}
where $v_A$ is the Alfv\'en velocity and $\delta$ is the rigidity dependence 
slope of the spatial diffusion coefficient \citep{1994ApJ...431..705S}.
The numerical package DRAGON is used to solve the propagation equation of GCRs 
\citep{2017JCAP...02..015E}. The propagation model parameters are listed in 
Table \ref{tab:diff_aniso}.

\begin{table*}[!htb]
	\centering
	\begin{center}
		\begin{tabular}{ccccccccccc}
			\toprule[1.5pt]
			& $D_0$  & $\delta_0$ & ${N}_m$ & $\xi$ & $n$ & $\eta$ & ${\cal R}_0$  & $v_A$  & $z_h$  \\
			&$[10^{28}$~cm$^2$/s] & $ $ & $ $ & $ $ & $ $ & $ $ & [GV] & [km/s] & [kpc] \\
			\hline
			w/ IBEX & 8.07  & ~0.45~  & ~0.9~ & ~0.1~ & ~3.5~ & ~0.05~ & 4 & 6 & 5\\
			w/o IBEX  & 18.9  & ~0.45~  & ~0.9~ & ~0.1~ & ~3.5~ & ~0.05~ & 4 & 6 & 5\\
			\bottomrule[1.5pt]
		\end{tabular}
	\end{center}
	\caption{Propagation parameters of the SDP model with/without the IBEX magnetic field.}
	\label{tab:diff_aniso}
\end{table*}

\subsection{The nearby source}
It is possible to have supernova explosions in the vicinity of the solar 
neighborhood not very long ago, which can accelerate high-energy particles 
and contribute to the locally observed GCRs. The observations of supernova 
remnants (SNRs) and pulsars do reveal a number of such candidates within 
e.g., 1 kpc from the Earth \cite{2020PhRvD.101l3023B}.
{Indirect evidence of nearby supernova explosions can also 
be obtained from the $^{60}$Fe observations\footnote{Note that, the explosion 
time of the supernova inferred from the $^{60}$Fe observations is about $2$ 
Myr, which is somehow different from the age of Geminga ($340$ kyr) as 
postulated to be the candidate local source in this analysis.}
\citep{2016Sci...352..677B,2016Natur.532...69W}. 
The imprints of nearby source(s) on the energy spectra of GCR nuclei, 
electrons and positrons have also been investigated  
\citep[e.g.,][]{2015PhRvL.115r1103K,2021JCAP...05..012Z}, which can
consistently explain the observational data. Particularly, the prominent 
positron excess can be explained by such a local supernova consistent 
with the $^{60}$Fe observations \citep{2015PhRvL.115r1103K}.}

While the identification of any specific SNR as the cosmic ray source is 
still challenging, here we take the Geminga as a {major 
contributor of the local sources for simplicity. Under the SDP scenario 
of GCR propagation, \citet{2022ApJ...930...82L} demonstrated that, 
among the observed local SNRs, only Geminga and Monogem could have 
important contributions to the GCR spectra. The contributions from other
sources got suppressed due to either their young ages or large distances.
This is mainly because of the much slower diffusion in the Galactic disk
compared with the conventional propagation model. {Note, however, a large amount of old sources should have enough time to transport throughout the Milky Way in the SDP scenario, giving a background GCR component as measured in the low-energy range.} It is still possible that 
both Geminga and Monogem contribute to the GCR spectra and anisotropies. 
Since their locations are similar, we expect that their impacts degenerate 
with each other. 
Furthermore, as illustrated in \citet{2018JCAP...07..051G},
the volume that particles emitted by a single source can be reduced 
significantly if the anisotropic diffusion is considered, and hence a
single source may dominate the local GCR fluxes.}

The Geminga pulsar locates at the 
direction of (R.A., Decl.)$=(6^h34^m,17^{\circ}46')$. Its distance is 
estimated to be about 250 pc, and the characteristic age is about 
$3.4\times10^5$ yrs \citep{2005AJ....129.1993M}. However, the Geminga 
pulsar has a remarkable proper motion with a transverse velocity of 
about 200 km~s$^{-1}$ \citep{2007Ap&SS.308..225F}, and its birth place 
is inferred to be close to the Orion association with (R.A., Decl.)
$=(5^h30^m, 10^{\circ}0')$ \citep{1994A&A...281L..41S,2007Ap&SS.308..225F}.
We therefore take its birth place as the direction of the Geminga SNR, 
and assume the same distance of 250 pc.

The propagation of particles from the nearby source is calculated using 
the Green's function method, assuming a spherical geometry with infinite 
boundary conditions. Assuming instantaneous injection from a point source, 
the GCR density as a function of space, rigidity, and time can be calculated as
\begin{equation}
\phi(r, {\cal R}, t) = \dfrac{q_{\rm inj}({\cal R})}{\left(\sqrt{2\pi} \sigma\right)^3} 
\exp \left(-\dfrac{r^2}{2\sigma^2} \right)~,
\end{equation}
where $q_{\rm inj}({\cal R})$ is the injection spectrum as a function of 
rigidity, $\sigma({\cal R}, t)=\sqrt{2D({\cal R})t}$ is the effective 
diffusion length within time $t$. The diffusion coefficient $D({\cal R})$ 
takes the solar system value of Eq. (1). The function form of 
$q_{\rm inj}({\cal R})$ is assumed to be power-law with an exponential 
cutoff. Table \ref{tab:para_aniso} gives the injection spectral 
parameters of both the background and the nearby source. 
{The total energy injected into cosmic rays from the local 
source is about $1.4\times 10^{50}$ erg for $E_k>$~GeV/n, which is consistent 
with the expectation from a typical supernova explosion.}

\begin{table*}
	\begin{center}
	\resizebox{1.\textwidth}{!}{
		\begin{tabular}{|c|ccc|ccc|ccc|ccc|}
			\hline
			& \multicolumn{3}{c|}{Bkg (w/ IBEX)} & \multicolumn{3}{c|}{Loc Src (w/ IBEX)} & \multicolumn{3}{c|}{Bkg (w/o IBEX)} & \multicolumn{3}{c|}{Loc Src (w/o IBEX)} \\
			\hline
			Element & Norm$^\dagger$ & ~~~$\nu$~~~  & ~~~$\mathcal R_{c}$~~~ & ~~~$q_0$~~~~~ & ~~~~~$\alpha$~~~ & ~~~${\cal R}'_c$~~~ & Norm$^\dagger$ & ~~~$\nu$~~~  & ~~~$\mathcal R_{c}$~~~ & ~~~$q_0$~~~~~ & ~~~~~$\alpha$~~~ & ~~~${\cal R}'_c$~~~ \\
			\hline
			& $[({\rm m}^2\cdot {\rm sr}\cdot {\rm s}\cdot {\rm GeV})^{-1}]$ & & [PV] & [GeV$^{-1}$] & &  [TV] & $[({\rm m}^2\cdot {\rm sr}\cdot {\rm s}\cdot {\rm GeV})^{-1}]$ & & [PV] & [GeV$^{-1}$] & &  [TV] \\
			\hline
			p   & $3.99\times 10^{-2}$    & 2.48   &  7  & $1.85\times 10^{52}$  & 2.32 & 30 & $3.80\times 10^{-2}$    & 2.47   &  7  & $3.70\times 10^{52}$  & 2.22 & 30 \\
			He & $2.45\times 10^{-3}$   & 2.40     &  7  & $9.01\times 10^{51}$  & 2.30  &  30 & $2.08\times 10^{-3}$   & 2.38     &  7  & $1.30\times 10^{52}$  & 2.15  &  30 \\
			C   & $8.93\times 10^{-5}$   & 2.42    &  7  & $3.20\times 10^{50}$    & 2.30 &  30 & $7.05\times 10^{-5}$   & 2.40    &  7  & $3.90\times 10^{50}$    & 2.15 &  30 \\
			N   & $1.31\times 10^{-5}$   & 2.46    &  7  & $4.50\times 10^{49}$  & 2.30 &   30 & $1.28\times 10^{-5}$   & 2.46    &  7  & $6.10\times 10^{49}$  & 2.15 &   30 \\
			O   & $1.05\times 10^{-4}$   & 2.42    &  7  & $3.48\times 10^{50}$ & 2.30  &   30 & $7.67\times 10^{-5}$   & 2.39    &  7  & $4.50\times 10^{50}$ & 2.15  &   30 \\
			Ne & $1.39\times 10^{-5}$   & 2.42   &  7  & $5.60\times 10^{49}$ & 2.30  &   30 & $1.06\times 10^{-5}$   & 2.38   &  7  & $5.02\times 10^{49}$ & 2.13  &   30 \\
			Mg & $1.97\times 10^{-5}$   & 2.44     &  7  & $5.60\times 10^{49}$ & 2.30  &   30 & $1.01\times 10^{-5}$   & 2.38     &  7  & $5.50\times 10^{49}$ & 2.13  &   30 \\
			Si & $1.69\times 10^{-5}$     & 2.45   &  7  & $5.60\times 10^{49}$ & 2.30  &   30 & $1.01\times 10^{-5}$     & 2.38   &  7  & $5.50\times 10^{49}$ & 2.13  &   30 \\
			Fe & $2.05\times 10^{-5}$    & 2.40    &  7  & $3.50\times 10^{49}$ & 2.30   &  30 & $1.12\times 10^{-5}$    & 2.37    &  7  & $4.21\times 10^{49}$ & 2.13   &  30 \\
			\hline
		\end{tabular}}\\
		$^\dagger${The normalization is set at kinetic energy per nucleon $E_{k} = 100$ GeV/n.}
	\end{center}
	\caption{Injection spectral parameters of the background and local source with/without 
		the alignment along the IBEX magnetic field.}
	\label{tab:para_aniso}
\end{table*}

\subsection{Alignment of GCR streaming along the magnetic field} 

Assuming there is an ordered, large-scale regular magnetic field, the 
propagation of GCRs may become anisotropic. In this case, the diffusion 
coefficient needs to be written as a tensor $D_{ij}$ \citep{2017JCAP...10..019C}
\begin{equation}
D_{ij} = D_{\perp}\delta_{ij} + (D_{\parallel}-D_{\perp})b_i b_j.
\end{equation}
Here, $D_{\parallel}$ and $D_{\perp}$ are the diffusion coefficients in 
parallel and perpendicular to the ordered magnetic field, $b_i$ = 
$B_i/{\textbf{B}}$ is the $i_{th}$ component of the unit vector of the 
magnetic field. 
We assume that $D_{\perp}=\varepsilon D_{\parallel}$ with $\varepsilon=0.15$. The perpendicular diffusion coefficient $D_{\perp}$ is adopted to be 
$D_{xx}(8.5~{\rm kpc},0,{\cal R})$ in the solar neighborhood. 


\section{Results}
The model considered in this work includes two source components, the 
background source diffusely distributed in the Milky Way which is assumed 
to be able to accelerate particles to energies beyond PeV, and a nearby 
source which mainly contributes to GCRs with energies $\lesssim100$ TeV. 
The propagation of GCRs in the Milky Way is spatially dependent, as 
suggested by the very-high-energy $\gamma$-ray observations of pulsars 
\citep{2017Sci...358..911A,2021PhRvL.126x1103A}. Specifically, particles 
diffuse relatively slowly in the Galactic plane and much faster in the 
halo, which is described by a two-halo approach (see Methods). The nearby 
source is assumed to be located in the outer Galaxy. We take the Geminga 
SNR, the birth place of Geminga pulsar after correcting 
its proper motion \citep{1994A&A...281L..41S,2007Ap&SS.308..225F}, as a 
benchmark illustration \citep{2019JCAP...10..010L}. However, it could be 
other source(s) with proper parameters. Since the local source is assumed 
to be nearby (with a distance of $0.25$ kpc), the diffusion coefficient 
is taken as the local value around the solar system. The other important 
effect in regulating the propagation of GCRs is the alignment of the GCR 
streaming along the large-scale magnetic field \citep{2016PhRvL.117o1103A}.
We take into account this effect via an anisotropic diffusion based on 
the IBEX magnetic field configuration. See the Methods for more details 
about the propagation model settings.

\begin{figure*}[!htb]
	\centering
	\includegraphics[width=0.49\textwidth]{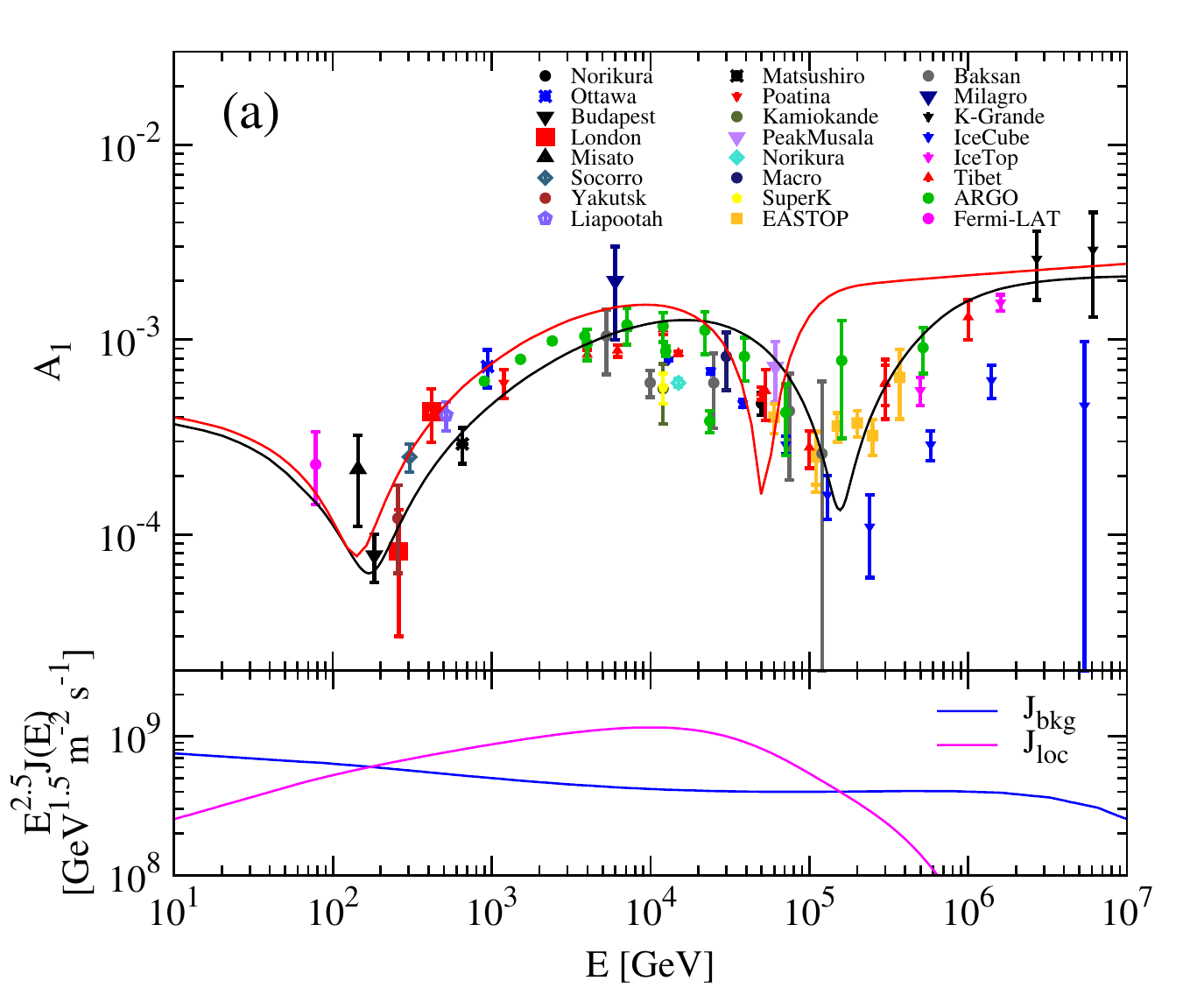}
	\includegraphics[width=0.49\textwidth]{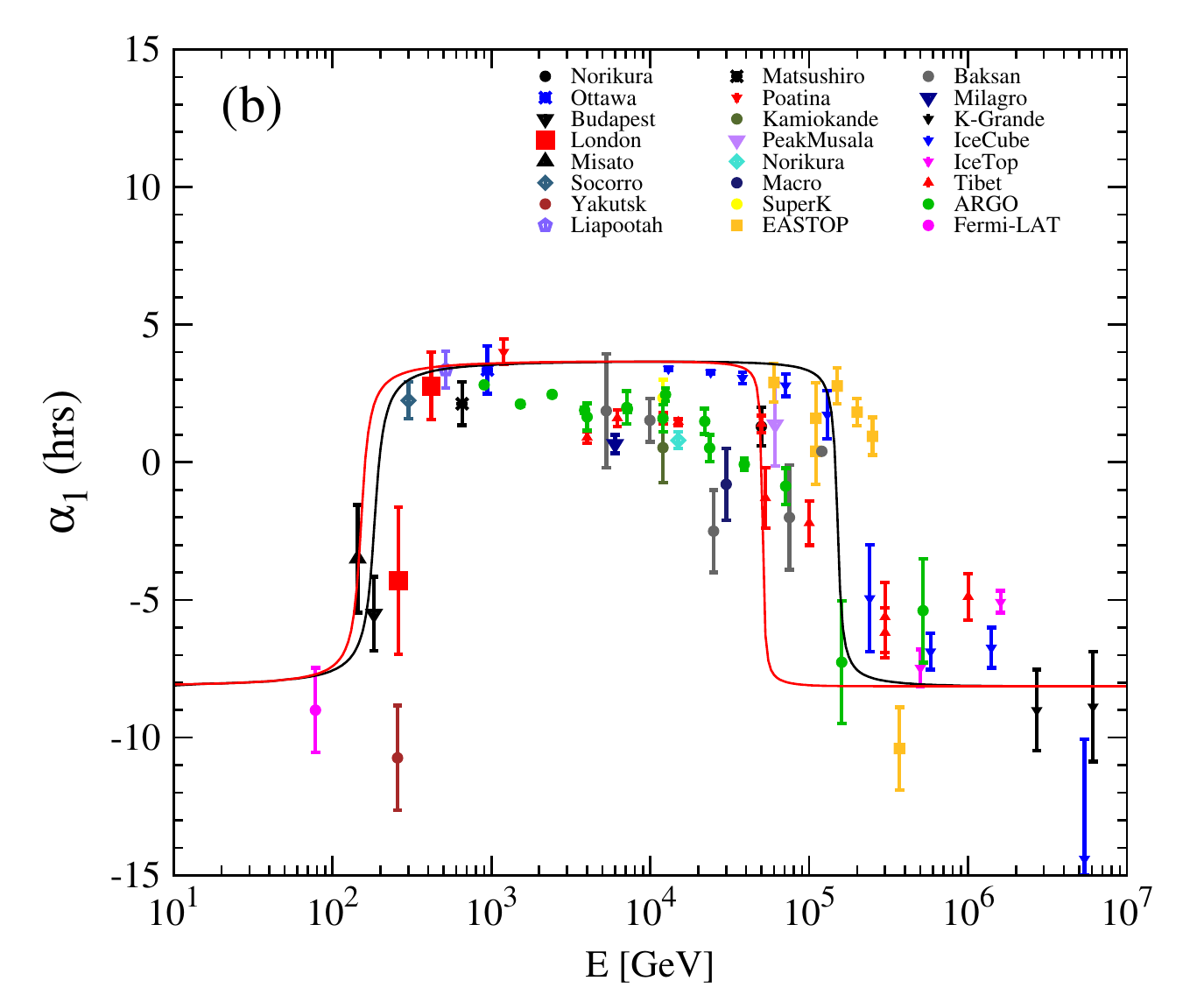}
	\includegraphics[width=0.49\textwidth]{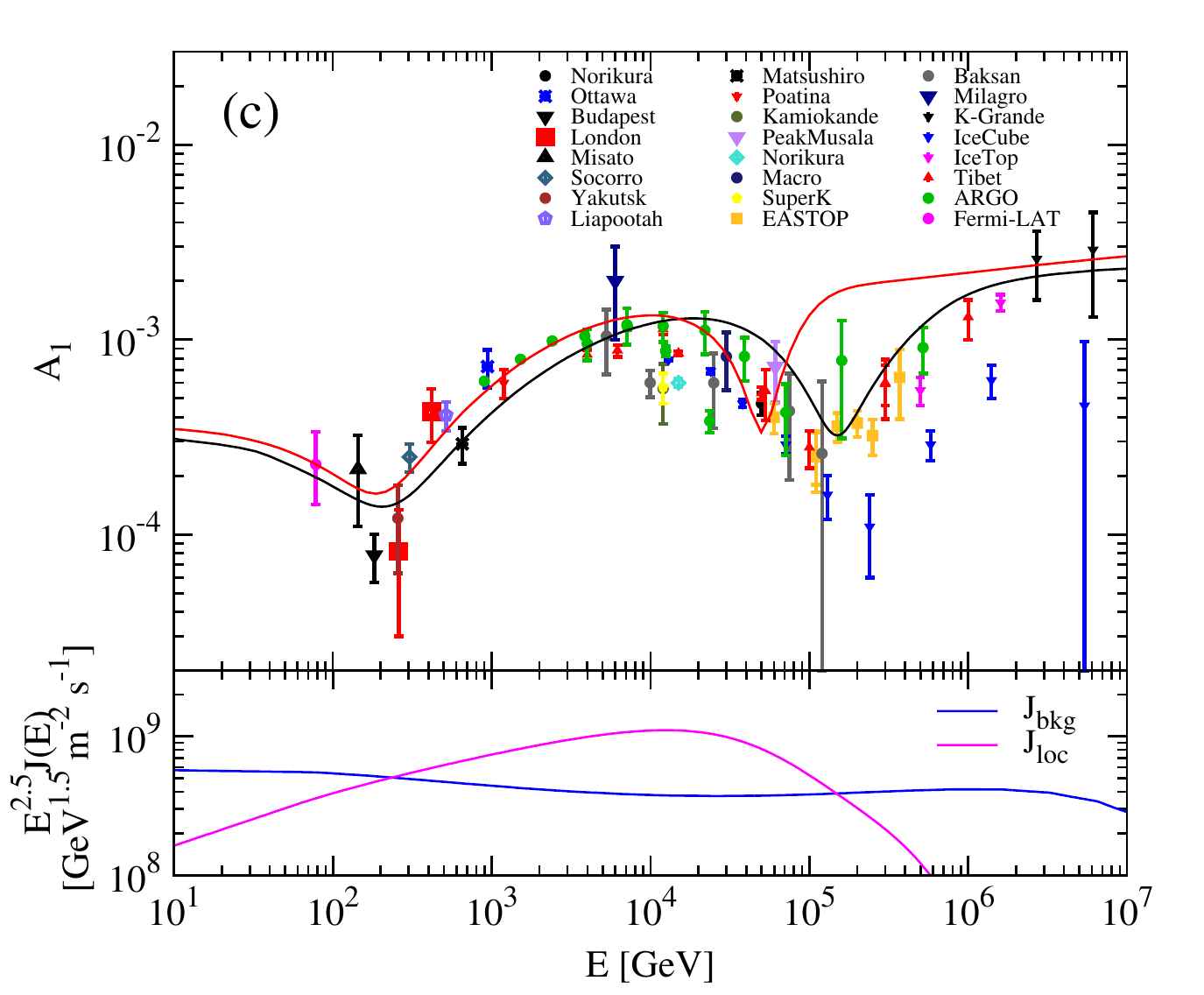}
	\includegraphics[width=0.49\textwidth]{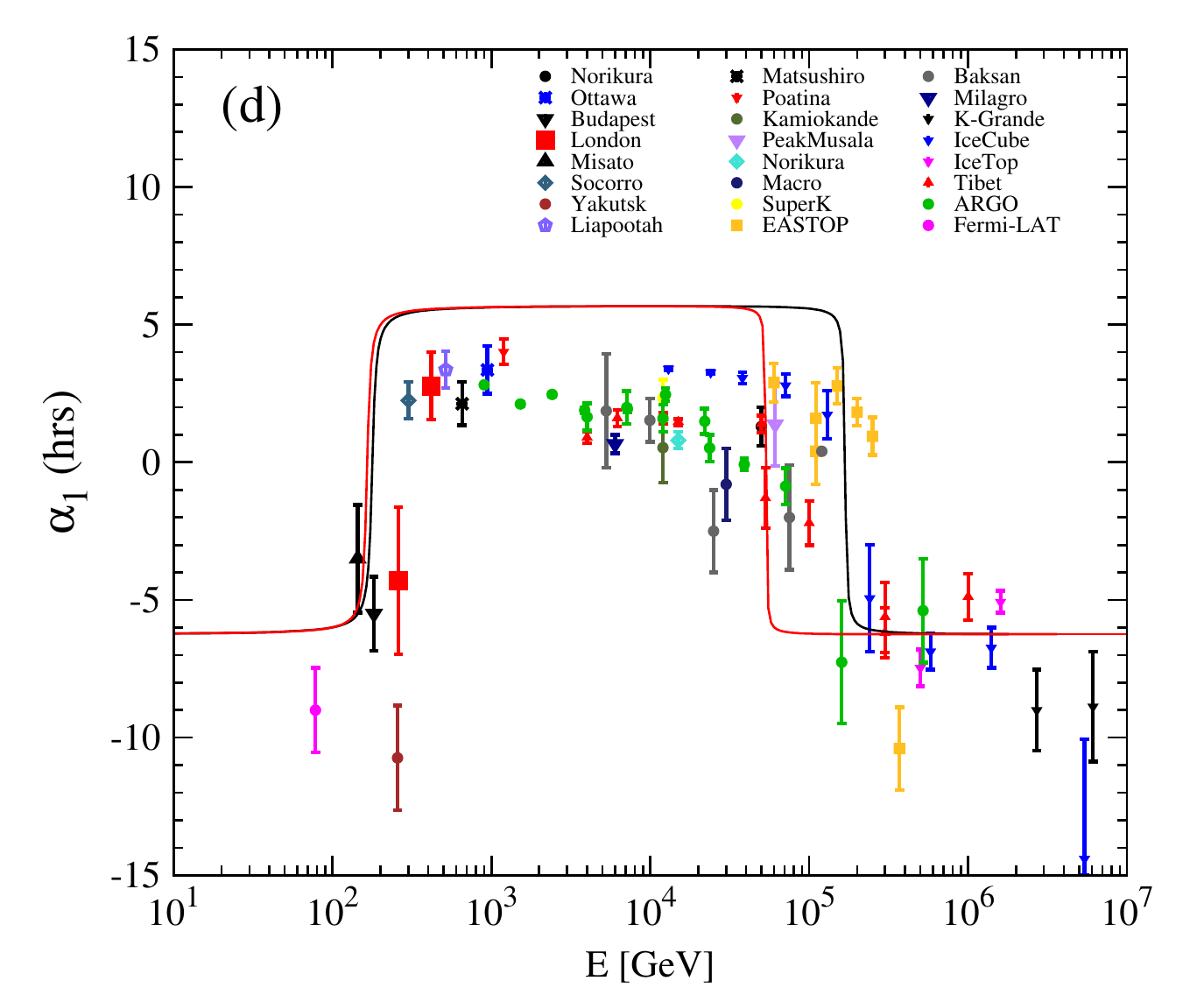}
	\caption{The model calculations of the amplitude and phase of the dipole 
		anisotropies, compared with the data. (a) and (b): The amplitude and phase 
		for the case with alignment along the IBEX magnetic field. (c) and (d): 
		The amplitude and phase for the case without alignment along the IBEX 
		magnetic field. In each panel, the black curve is for all GCR species, 
		and the red one is for protons. The bottom sub-panels of (a) and (c) show 
		the streamings of all species from the background source (blue curve) and 
		the nearby source (magenta curve).}
	\label{fig:aniso}
\end{figure*}

Fig. \ref{fig:aniso} shows the calculated amplitudes and phases of the 
dipole anisotropies of GCRs in a wide energy range from 10 GeV to $>$PeV, 
compared with the measurements 
\citep{1973ICRC....2.1058S,1975ICRC....2..586G,1981ICRC...10..246B,
	1981ICRC....2..146A,1983ICRC....3..383T,1985P&SS...33..395N,1985P&SS...33.1069S,
	1987ICRC....2...22A,1989NCimC..12..695N,1995ICRC....4..639M,1995ICRC....4..648M,
	1995ICRC....4..635F,1995ICRC....2..800A,1996ApJ...470..501A,1997PhRvD..56...23M,
	2003PhRvD..67d2002A,2005ApJ...626L..29A,2007PhRvD..75f2003G,2009ApJ...692L.130A,
	2009NuPhS.196..179A,2009ApJ...698.2121A,2010ApJ...718L.194A,2012ApJ...746...33A,
	2013ApJ...765...55A,2015ApJ...809...90B,2017ApJ...836..153A,2018ApJ...861...93B,
	2019ApJ...883...33A}. 
The measurements made by the Fermi satellite are for protons 
\citep{2019ApJ...883...33A}, while the other measurements by underground 
muon detectors and ground-based air shower detectors are for all particle 
species in GCRs. Therefore, in each plot we show the results for protons 
(red) and all species (black) individually. As can be expected from the 
model, two phase reversals should be observed in the anisotropies, at 
energies when the GCR streaming from the nearby source becomes dominant 
($E\sim 100$ GeV) and sub-dominant ($E\sim100$ TeV) compared with the 
background streaming. The competition of the two particle streamings
leaves also imprints on the amplitudes, characterized by two dip-like 
structures at corresponding energies. These features are largely 
consistent with the measurements.

The local large-scale magnetic field may have a sizeable impact on the 
observed phases of anisotropies. Panels (c) and (d) of Fig. \ref{fig:aniso} 
give the results without the alignment along the IBEX magnetic field. 
The dips of the amplitudes become less distinct, because the two 
streamings from the background and the nearby source are not in the 
exactly opposite directions. After the projection along the magnetic field,
they tend to cancel with each other at specific energies, resulting in 
deeper dips. Even bigger effects can be seen for the phases. Without 
considering the alignment effect due to the magnetic field, the 
anisotropy phases are determined solely by the source distributions. 
Therefore, for both low ($\lesssim 100$ GeV) and high ($\gtrsim 100$ TeV) 
energies, the excesses should point to the GC direction with R.A.$\sim -6$ 
hr. For intermediate energies the excesses point to the direction of
the assumed nearby source with R.A.$\sim 5.5$ hr. The anisotropic 
diffusion results in an alignment along the IBEX magnetic field, and 
the phases change to $\sim -8$ hr for low and high energies and $\sim 4$ 
hr for intermediate energies, which are more consistent with the data.

\begin{figure}[htb]
	\centering
	\includegraphics[width=0.55\textwidth]{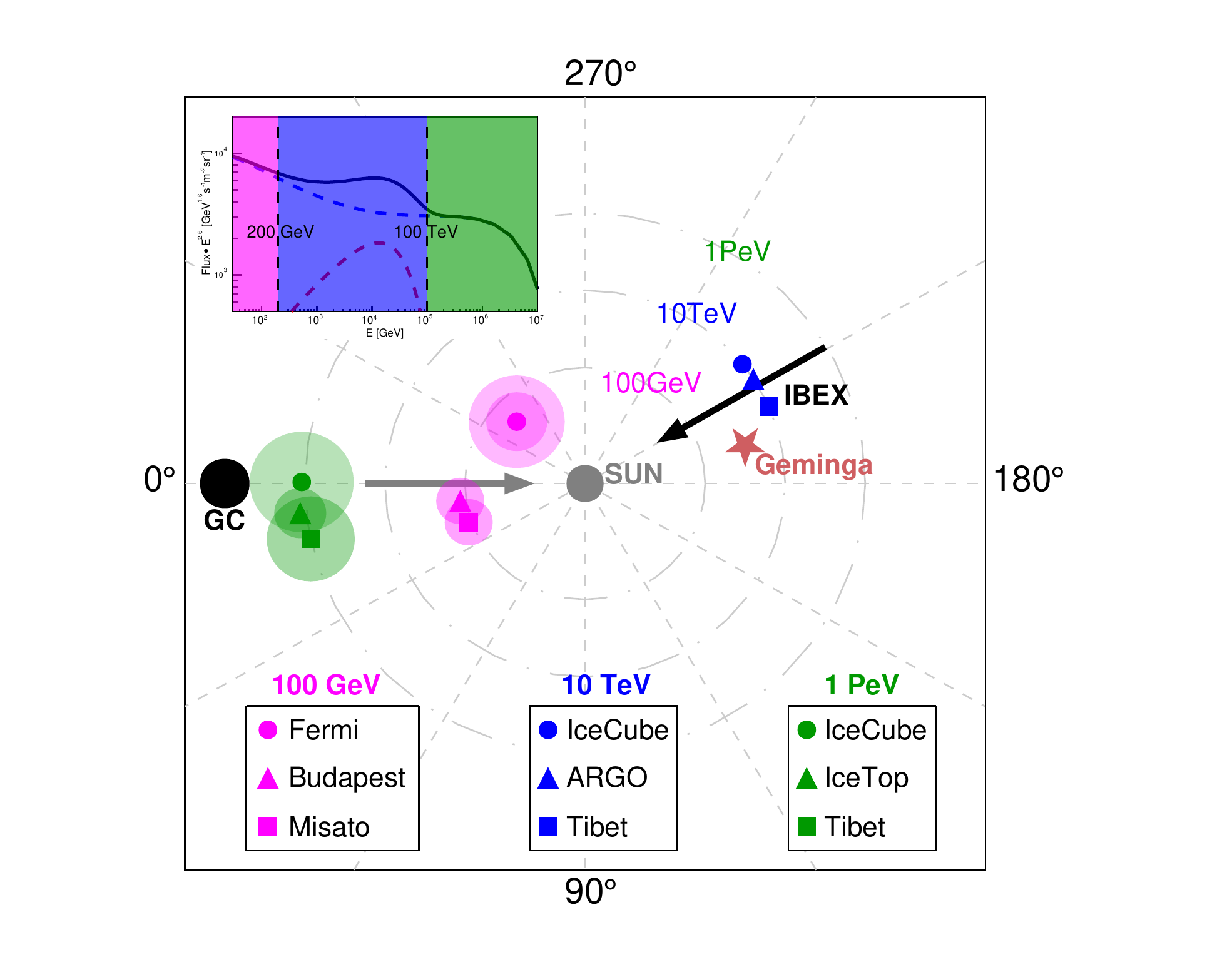}
	\caption{The cartoon illustration of the co-evolution of spectra and 
		anisotropies. This plot is a top-down view of the Milky Way from the 
		Galactic north pole to the south pole, centering at the Sun. The black 
		(grey) arrow shows the GCR streaming from the nearby (background) source. 
		The dots, squares, and triangles show the observed excess directions of 
		the anisotropies, with typical energies of 100 GeV (magenta), 10 TeV 
		(blue), and 1 PeV (green), respectively. The directions of the GC, 
		the Geminga SNR, and the IBEX magnetic field are also labelled.
		The inserted plot illustrates the structures of GCR spectra in this 
		model, with three energy bands being shown by three colors.}
	\label{fig:cartoon}
\end{figure}


\section{CONCLUSION}
In this work, we propose a simple two-component model to simultaneously 
explain the spectral features of GCR nuclei (including the hardening aroung 
$200$ GV and softening around $10$ TV), the amplitudes and two phase 
reversals of the dipole anisotropies. The energy spectra and anisotropies 
can thus be understood as the scalar and vector parts of this model, and 
a unified understanding of their complicated observational properties 
can be achieved in this two-component scenario. The algebraic sum of the 
fluxes of the background sources and the nearby source results in 
spectral hardenings around $O(200)$ GV and softenings around $O(10)$ 
TV rigidities, while the vector sum of the streamings gives the phase
reversals and amplitude variations of the anisotropies. 
Fig. \ref{fig:cartoon} shows a schematic illustration of the co-evolution 
of the energy spectra and anisotropies of GCRs. We note that, the 
anisotropy phases are more sensitive to reveal different components
of GCRs than the spectra, due to the vector sum nature.

While this model can explain the overall observed results of the 
spectra and large-scale anisotropies, we note that the 
two-dimensional anisotropies may not be perfectly reproduced by this 
simple model. As expected by the model, for $E\lesssim100$ GeV, the 
dipole anisotropies should be dominated by the background source 
component, which should be aligned to the (anti-)direction of the IBEX 
magnetic field (R.A.$\sim -9$ hr, Decl.$\sim21^{\circ}$). However, the 
Fermi measurements showed that the two-dimensional anisotropy excess 
points to R.A.$=-9.7\pm1.5$ hr, Decl.$=-51^{\circ}\pm21^{\circ}$. 
The R.A. of the dipole anisotropy is well consistent with the model 
prediction, while the declination shows a deviation by about $3\sigma$. 
The significance of the Fermi observation is not high enough. Better 
measurements by future experiments like DAMPE \citep{2017APh....95....6C}
and HERD \citep{2014SPIE.9144E..0XZ} are necessary to further understand 
whether there is a discrepancy between the model and the data. The 
two-dimension anisotropy measurements for groundbased experiments are 
subject to difficulties of the absolute efficiency calibration, and are 
thus difficult for precise model tests.

The measurements of a few underground muon detectors were ambiguous. 
Measurements at different time showed variations of the anisotropy patterns 
for the muon detectors, such as London and Socorro 
\citep{1972Natur.235...25S,1974JGR....79.3695S}. In Fig. \ref{fig:aniso} we show 
the average sidereal component of the anisotropies measured by the muon detectors. 
Several models were proposed to understand the origin of such variations, such as 
the Jupiter's magnetosphere \citep{1974JGR....79.3695S}, the polarity reversal of 
the solar magnetic field \citep{1975ICRC....4.1453C}, or a variable asymmetric 
anisotropy component in the northern hemisphere which might be of solar origin 
\citep{1972RISRJ..26...31N,1983NCimC...6..550N}. Observational evidence against 
those proposals also appeared \citep{1974PASA....2..293J,1984NCimC...7..379N,1985P&SS...33..395N,1985P&SS...33.1069S}, 
and no consensus has been reached yet. 
We note that, in our model, the anisotropy patterns in the $O(100)$ GeV energy range 
are sensitive to the energy scales of the experiments, as shown in Fig. \ref{fig:aniso}. 
The median energy of muon detectors might vary with time, due to the instrument
calibration and/or the solar modulation. In addition, the energy resolution of
those detectors is relatively poor. It is possible that the variations of the
anistropy phases were due to the variations of median energies of different samples.
Future direct detection experiments may test this hypothesis with more precise 
measurements in the low-energy band.

The second phase reversal around 100 TeV is also not well measured by 
current experiments. It is difficult to see at which energy the phase 
starts to reverse and how it changes from one phase to the other. 
Current indirect detection experiments are subject to large uncertainties 
of the composition discrimination, and the anisotropies are for mixed 
GCRs of all nuclei species. For energies above 100 TeV, it is also not 
clear whether the GCR anisotropy phase is closer to the GC or still 
aligned along the local magnetic field. We expect that improved 
measurements of the anisotropies of various mass groups by e.g., 
LHAASO \citep{2019arXiv190502773C} can be very helpful in testing our 
model and understanding the origin of GCR anisotropies.

The idea of this work may further apply to the energy spectra and 
anisotropies of cosmic rays at even higher energies, where the transition 
from Galactic to extragalactic origins occurs. The Pierre Auger Observatory 
reported the measurements of the anisotropies above 8 EeV, with direction 
of R.A.$\sim6.7\pm0.7$ hr and Decl.$=-24^{+12\circ}_{-13}$, which differs 
significantly from the GC but may be consistent with the infrared galaxy 
distribution after correcting the Galactic magnetic field effect on the 
anisotropies \citep{2017Sci...357.1266P}. Improved measurements of how 
the anisotropies evolve between PeV and EeV energies may be used to 
diagnose the Galactic-extragalactic transition of cosmic rays.

\section*{Acknowledgements}
This work is supported by the National Key Research and Development Program 
(No. 2018YFA0404202), the National Natural Science Foundation of China 
(Nos. 12220101003, 12275279), and the Project for Young Scientists in Basic 
Research of Chinese Academy of Sciences (No. YSBR-061).

\bibliographystyle{apj}
\bibliography{refs}

\end{document}